\documentclass[aps,preprint,amsmath,amssymb]{revtex4}

\usepackage{graphicx}
\begin{document}

\title{Anomalous quartic $ZZ\gamma\gamma$ couplings in $\gamma p$ collision at the LHC}

\author{\.{I}. \c{S}ahin}
\email[]{inancsahin@karaelmas.edu.tr}
 \affiliation{Department of Physics, Bulent Ecevit University, 67100 Zonguldak, Turkey}

\author{B. \c{S}ahin}
\email[]{bsahin@karaelmas.edu.tr} \affiliation{Department of
Physics, Bulent Ecevit University, 67100 Zonguldak, Turkey}

\begin{abstract}
We investigate the constraints on the anomalous quartic
$ZZ\gamma\gamma$ couplings through the process $p p\to p\gamma p\to
p \gamma q Z X$ at the LHC. Taking into consideration various
forward detector acceptances and integrated LHC luminosities, we
find 95\% confidence level bounds on the anomalous coupling
parameters. We show that the bounds on these couplings are at the
order of $10^{-6}\;\textmd{GeV}^{-2}$ which are about four orders of
magnitude more restricted with respect to current experimental
bounds.

\end{abstract}

\maketitle

\section{Introduction}

Precision measurements of gauge boson self-interactions at the LHC
will be the crucial test of the $SU_L(2)\times U_Y(1)$ gauge
structure of the standard model (SM). Any deviation of the couplings
from the SM expectations would indicate the existence of new
physics. It is very common to investigate the new physics via
effective Lagrangian approach. The theoretical basis of such an
approach rely on the assumption that at higher energies beyond the
SM, there is a more fundamental theory which reduces to the SM at
lower energies.  Hence, SM is assumed to be an effective low-energy
theory in which heavy fields have been integrated out. Such a
procedure is quite general and independent of the details of the
model. For this reason this approach is sometimes called model
independent analysis.

In this paper we have analyzed genuine quartic $ZZ\gamma\gamma$
couplings via single Z boson production in a $\gamma$-proton
collision at the LHC. Genuine quartic couplings have different
origins than anomalous trilinear couplings. They arise from
effective operators that do not induce any trilinear gauge boson
coupling. Hence, genuine quartic couplings are free from constraints
on trilinear couplings. Imposing custodial $SU(2)_{Weak}$ symmetry
and local $U(1)_{em}$ symmetry, C and P conserving dimension 6
effective lagrangian for $ZZ\gamma\gamma$ couplings are given by
\cite{boudjema1,boudjema2,eboli1},
\begin{eqnarray}
{\cal L}&&={\cal L}_0+{\cal L}_c
\end{eqnarray}
\begin{eqnarray}
{\cal
L}_{0}&&=\frac{-\pi\alpha}{4\Lambda^{2}}a_{0}F_{\mu\nu}F^{\mu\nu}
W_{\alpha}^{(i)}W^{(i) \alpha}
\end{eqnarray}
\begin{eqnarray}
{\cal
L}_{c}&&=\frac{-\pi\alpha}{4\Lambda^{2}}a_{c}F_{\mu\alpha}F^{\mu\beta}
W^{(i) \alpha}W_{\beta}^{(i)}
\end{eqnarray}
where $W^{(i)}$ is the $SU(2)_{Weak}$ triplet, and $F_{\mu\nu}$ is
the electromagnetic field strength. $a_0$ and $a_c$ are the
dimensionless anomalous coupling constants and $\Lambda$ is the
scale of new physics. The vertex functions generated from the
effective lagrangians (2) and (3) are given respectively by
\cite{eboli1}
\begin{eqnarray}
i\frac{2\pi\alpha}{\cos^{2}\theta_{W}\Lambda^{2}}a_{0}g_{\mu\nu}\left[g_{\alpha\beta}
(p_{1}.p_{2})-p_{2 \alpha}p_{1 \beta}\right]
\end{eqnarray}
\begin{eqnarray}
i\frac{\pi\alpha}{2\cos^{2}\theta_{W}\Lambda^{2}}a_{c}\left[(p_{1}.p_{2})(g_{\mu\alpha}
g_{\nu\beta}+g_{\mu\beta}g_{\alpha\nu})+g_{\alpha\beta}(p_{1\mu}p_{2\nu}
+p_{2\mu}p_{1\nu})\right. \nonumber \\
\left. -p_{1\beta}(g_{\alpha\mu}p_{2\nu}+g_{\alpha\nu}p_{2\mu})
-p_{2\alpha}(g_{\beta\mu}p_{1\nu}+g_{\beta\nu}p_{1\mu})\right]
\end{eqnarray}
where $p_{1}$ and $p_{2}$ are the momenta of photons and for a
convention, we assume that all the momenta are incoming to the
vertex.

The current best limits on $ZZ\gamma\gamma$ couplings are provided
by the OPAL Collaboration. These are
\begin{eqnarray}
 -0.007\; \textmd{GeV}^{-2} <
\frac{a_{0}}{\Lambda^{2}} < 0.023\; \textmd{GeV}^{-2} \\
-0.029\; \textmd{GeV}^{-2} < \frac{a_{c}}{\Lambda^{2}} < 0.029\;
\textmd{GeV}^{-2}
\end{eqnarray}
at 95\% C.L. \cite{Abbiendi:2004bf}.

Studying photon-induced reactions in a hadron collider is not a very
new phenomena. The reactions such as $p \bar p\to p \gamma \gamma
\bar p\to p e^+ e^- \bar p$ \cite{cdf1,cdf4}, $p \bar p\to p \gamma
\gamma \bar p\to p\; \mu^+ \mu^- \bar p$ \cite{cdf3,cdf4}, $p \bar
p\to p \gamma \bar p\to p\; J/\psi\;(\psi(2S)) \bar p$ \cite{cdf3}
were verified experimentally by the CDF collaboration at the
Fermilab Tevatron. These results raise interest on the potential of
LHC as a photon-photon and photon-proton collider. Probing new
physics via photon-induced reactions at the LHC is common in the
literature. Phenomenological studies involve: supersymmetry,
extradimensions, unparticle physics, gauge boson self-interactions,
neutrino electromagnetic properties etc.
\cite{fd11,fd12,fd13,lhc1,lhc2,lhc3,lhc4,lhc5,lhc6,lhc7,lhc8,lhc9,lhc10,lhc11,lhc12}.

\section{Equivalent photon approximation and cross sections}

\begin{figure}
\includegraphics[scale=1]{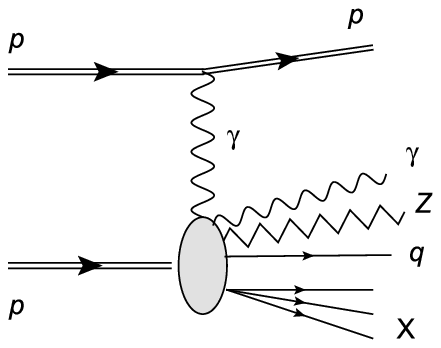}
\caption{Schematic diagram for the process $p p\to p\gamma p\to p
\gamma q Z X$. \label{fig1}}
\end{figure}

A quasireal photon emitted from one proton beam can interact with
the quarks of the other proton and the subprocess $\gamma q\to
\gamma q Z$ can occur at the LHC. Emitted quasireal photons are
described by equivalent photon approximation (EPA)
\cite{budnev,Baur,fd18}. Their virtuality is very low and it is a
good approximation to assume that they are on-mass-shell. Therefore
to some extent it is possible to study $\gamma$-proton collision at
the LHC. A schematic diagram describing this process is given in
Fig.\ref{fig1}.

Any process in a $\gamma$-proton collision can be discerned from
pure deep inelastic scattering processes by means of two
experimental signatures \cite{rouby}: First signature is the forward
large-rapidity gap. Quasireal photons have a low virtuality and
scattered with small angles from the beam pipe. Since the transverse
momentum carried by a quasireal photon is small, photon emitting
intact protons should also be scattered with small angles and exit
the central detector without being detected. This causes a decrease
in the energy deposit in the corresponding forward region compared
to the case in which the proton remnants are detected by the
calorimeters. As a result of this, one of the forward regions of the
central detector has a significant lack of energy. This defines the
forward large-rapidity gap and usual pp deep inelastic processes can
be rejected by applying a selection cut on this quantity. Second
experimental signature is provided by the forward detectors. Forward
detectors are capable to detect particles with a large
pseudorapidity. When a photon emitting intact proton is scattered
with a large pseudorapidity, it exceeds the pseudorapidity coverage
of the central detectors. The detection of this intact proton by the
forward detectors provides a distinctive signal for the
$\gamma$-proton collision.

$\gamma\gamma$ collision can also be studied in the framework of
EPA. In fact, these two-photon processes are much more studied at
the LHC. Two-photon processes provide a more clean environment with
respect to $\gamma$-proton processes due to absence of the remnants
of both proton beams. Moreover two-photon processes are generally
electroweak in nature and they are mostly free from backgrounds
coming from strong interactions. It is more probable for strong
interactions that take part in a $\gamma$-proton process rather than
in a two-photon process. Hence, two-photon processes generally have
less backgrounds compared to $\gamma$-proton processes. On the other
hand, for $\gamma$-proton processes energy reach and effective
luminosity are much higher than for two-photon processes
\cite{rouby,deFavereaudeJeneret:2009db}. This feature might be
important in probing new physics especially when the energy
dependences of the anomalous cross sections are very high. For
instance, anomalous $ZZ\gamma\gamma$ couplings are described by
effective lagrangians (2) and (3) which have an energy dimension of
6. Therefore anomalous cross section containing the $ZZ\gamma\gamma$
vertex has a higher momentum dependence than the SM cross section.
From a simple dimensional analysis we deduce that its momentum
dependence is higher up to a factor of $p^4$ than the SM cross
section where $p$ is the momentum incoming (or outgoing) to the
vertex. Thus, new physics contribution to the cross section rapidly
increases when the center-of-mass energy increases and the processes
which have a higher energy reach are expected to have a high
sensitivity to new physics. Another factor which has to be
considered in a $\gamma$-proton or a $\gamma\gamma$ collision is the
survival probability. It is basically defined as the probability of
the scattered protons not to dissociate due to the secondary soft
interactions \cite{deFavereaudeJeneret:2009db}. In the
$\gamma$-proton collision with quasireal photons virtuality of
photons is very small. In the EPA that we have considered typical
photon virtuality is $\langle Q^2\rangle \approx0.01 GeV^2$
\cite{fd18}. Therefore proton impact parameter is much bigger than
the range of strong interactions and proton survival probability is
expected to be large. On the other hand, the survival probability
for a $\gamma$-proton process is usually smaller than for a
two-photon process \cite{deFavereaudeJeneret:2009db}.


ATLAS and CMS collaborations have a program of forward physics with
extra detectors located at distances of 220m and 420m from the
interaction point \cite{royon,albrow}. These forward detectors have
a capability to detect intact scattered protons with momentum
fraction loss in the interval $\xi_{min} <\xi < \xi_{max}$ which is
called the acceptance of the forward detectors. The acceptance
proposed by the ATLAS Forward Physics (AFP) Collaboration is $0.0015
<\xi < 0.15$ \cite{royon,albrow}.  The acceptance of the forward
detectors in CMS is similar. There are also other scenarios with
different acceptances. CMS-TOTEM forward detector scenario spans
$0.0015 <\xi < 0.5$ \cite{avati,fd11}.

\begin{figure}
\includegraphics[scale=1]{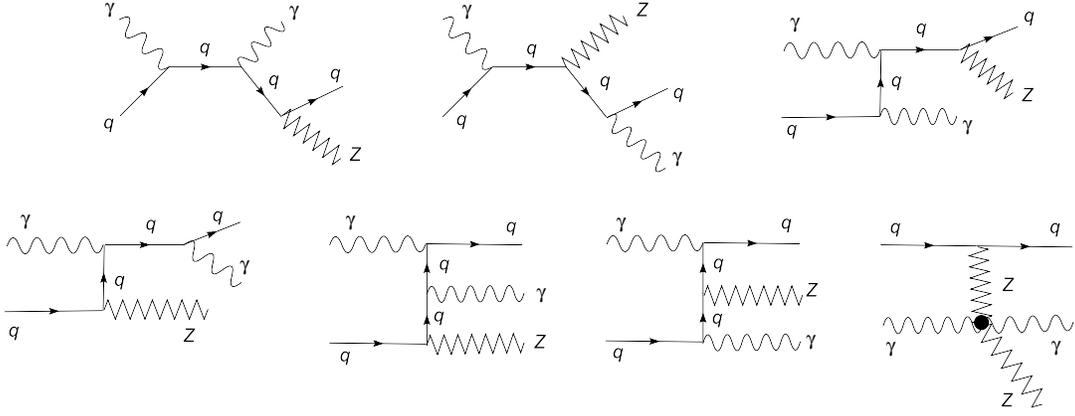}
\caption{Tree-level Feynman diagrams for the subprocess $\gamma q\to
\gamma q Z$ ($q=u,d,c,s,b,\bar u,\bar d, \bar c, \bar s, \bar b$).
\label{fig2}}
\end{figure}

The process $\gamma q\to \gamma q Z$ takes part as a subprocess in
the main reaction $p p\to p\gamma p\to p \gamma q Z X$. We consider
totaly 10 subprocesses for different type of quarks and anti-quarks:
\begin{eqnarray}
\label{subprocesses} &&\text{(i)}\;\;\gamma u \to \gamma u Z
\;\;\;\;\;\;\;\;\;\;\;\;\text{(vi)}\;\;\gamma \bar u \to \gamma \bar
u Z \nonumber
\\ &&\text{(ii)}\;\;\gamma d \to \gamma d Z
\;\;\;\;\;\;\;\;\;\;\;\;\text{(vii)}\;\;\gamma \bar d \to \gamma
\bar d Z\nonumber
\\&&\text{(iii)}\;\;\gamma c \to \gamma c Z
\;\;\;\;\;\;\;\;\;\;\text{(viii)}\;\;\gamma \bar c \to \gamma \bar c Z \\
&&\text{(iv)}\;\;\gamma s \to \gamma s Z
\;\;\;\;\;\;\;\;\;\;\;\text{(ix)}\;\;\gamma \bar s \to \gamma \bar s
Z \nonumber \\&&\text{(v)}\;\;\gamma b \to \gamma b Z
\;\;\;\;\;\;\;\;\;\;\;\;\;\text{(x)}\;\;\gamma \bar b \to \gamma
\bar b Z \nonumber
\end{eqnarray}
Each of the subprocesses is described by seven tree-level
diagrams(Fig.\ref{fig2}). We see from Fig.\ref{fig2} that one of
them contains anomalous $ZZ\gamma\gamma$ vertex and others are SM
contributions. The cross section for the main process $p p\to
p\gamma p\to p \gamma q Z X$ can be obtained by integrating the
cross sections for the subprocesses over the photon and quark
spectra:
\begin{eqnarray}
\label{crossection} \sigma\left(p p\to p\gamma p\to p \gamma q Z
X\right)=\sum_q\int_{x_{1\; min}}^{x_{1\;max}} {dx_1 }\int_{0}^{1}
{dx_2}\left(\frac{dN_\gamma}{dx_1}\right)\left(\frac{dN_q}{dx_2}\right)
\hat{\sigma}_{\gamma q\to \gamma q Z}(\hat s)\nonumber \\
=\sum_q\int_{\frac{M_{inv}}{\sqrt{s}}}^{\sqrt{\xi_{max}}}
{dz\;2z}\int_{MAX(z^2,\xi_{min})}^{\xi_{max}}
{\frac{dx_1}{x_1}}\left(\frac{dN_\gamma}{dx_1}\right)N_q(z^2/x_1)\;
\hat{\sigma}_{\gamma q\to \gamma q Z}(z^2s)
\end{eqnarray}
where $x_1$ is the fraction which represents the ratio between the
scattered equivalent photon and initial proton energy and $x_2$ is
the momentum fraction of the proton's momentum carried by the quark.
$\frac{dN_\gamma}{dx_1}$ is the equivalent photon spectrum (see
Appendix) and $\frac{dN_q}{dx_2}$ is the quark distribution function
of the proton.  The summations in (\ref{crossection}) are performed
over subprocesses in (\ref{subprocesses}). The second integral in
(\ref{crossection}) is obtained by transforming the differentials
$dx_1dx_2$ into $dzdx_1$ with a Jacobian determinant $2z/x_1$ where
$z=\sqrt{x_1x_2}\simeq \sqrt{\hat s/s}$. $M_{inv}$ is the total mass
of the final particles of the subprocess $\gamma q\to \gamma q Z$.
$N_q(z^2/x_1)$ is $\frac{dN_q}{dx_2}$ evaluated at $x_2=z^2/x_1$. At
high energies greater than proton mass, $\xi\simeq x_1$ holds.
Therefore it is a good approximation to assume that
$x_{1\;max}=\xi_{max}$ and $x_{1\;min}=\xi_{min}$ in the first
integral in (\ref{crossection}). During calculations, the virtuality
of the quark is taken to be ${Q}^2={m_Z}^2$ where $m_Z$ is the mass
of the Z boson. In our calculations parton distribution functions of
Martin, Stirling, Thorne and Watt \cite{pdf} have been used.

\begin{figure}
\includegraphics[scale=1]{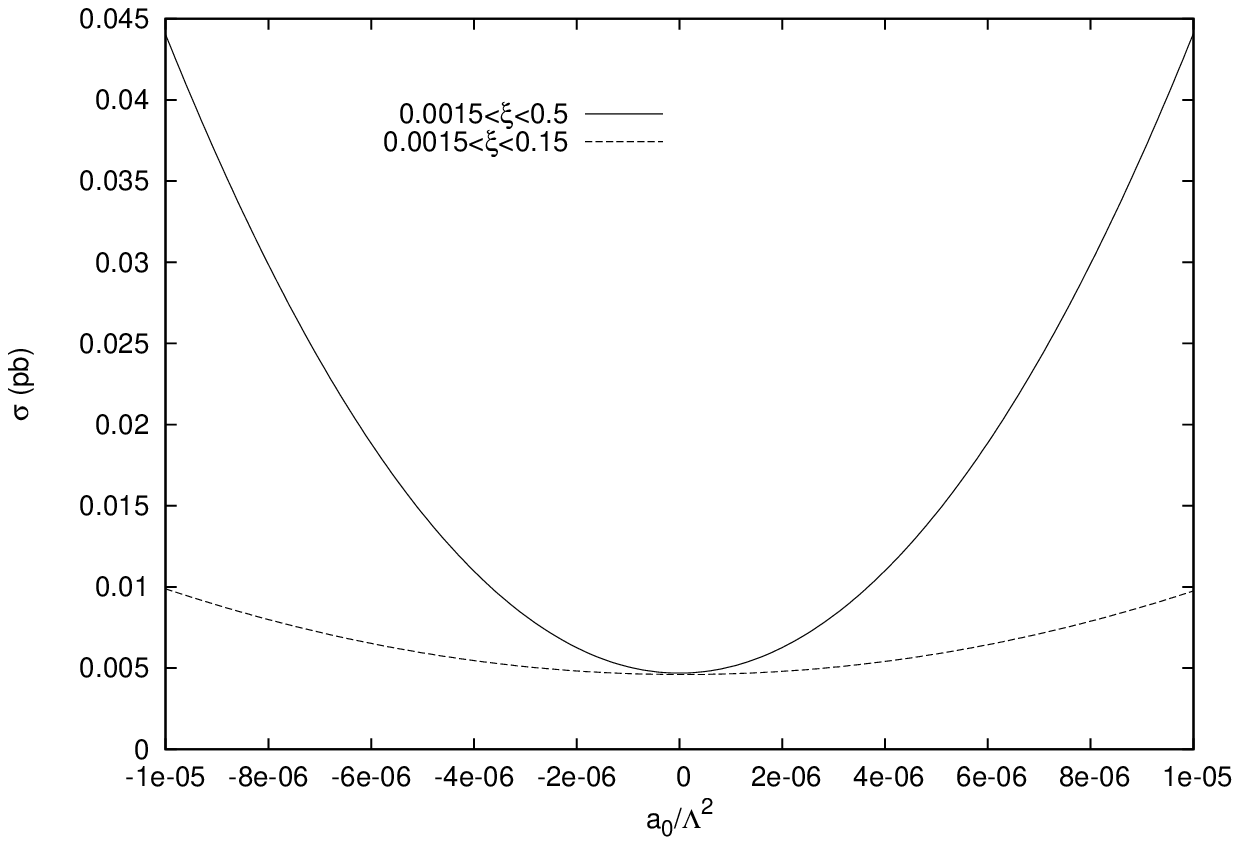}
\caption{Total cross section of $p p\to p\gamma p\to p \gamma q Z X$
as a function of anomalous coupling $\frac {a_{0}}{\Lambda^2}$ for
two different forward detector acceptances stated in the figure. The
center-of-mass energy of the proton-proton system is taken to be
$\sqrt s=14$ TeV. \label{fig3}}
\end{figure}

\begin{figure}
\includegraphics[scale=1]{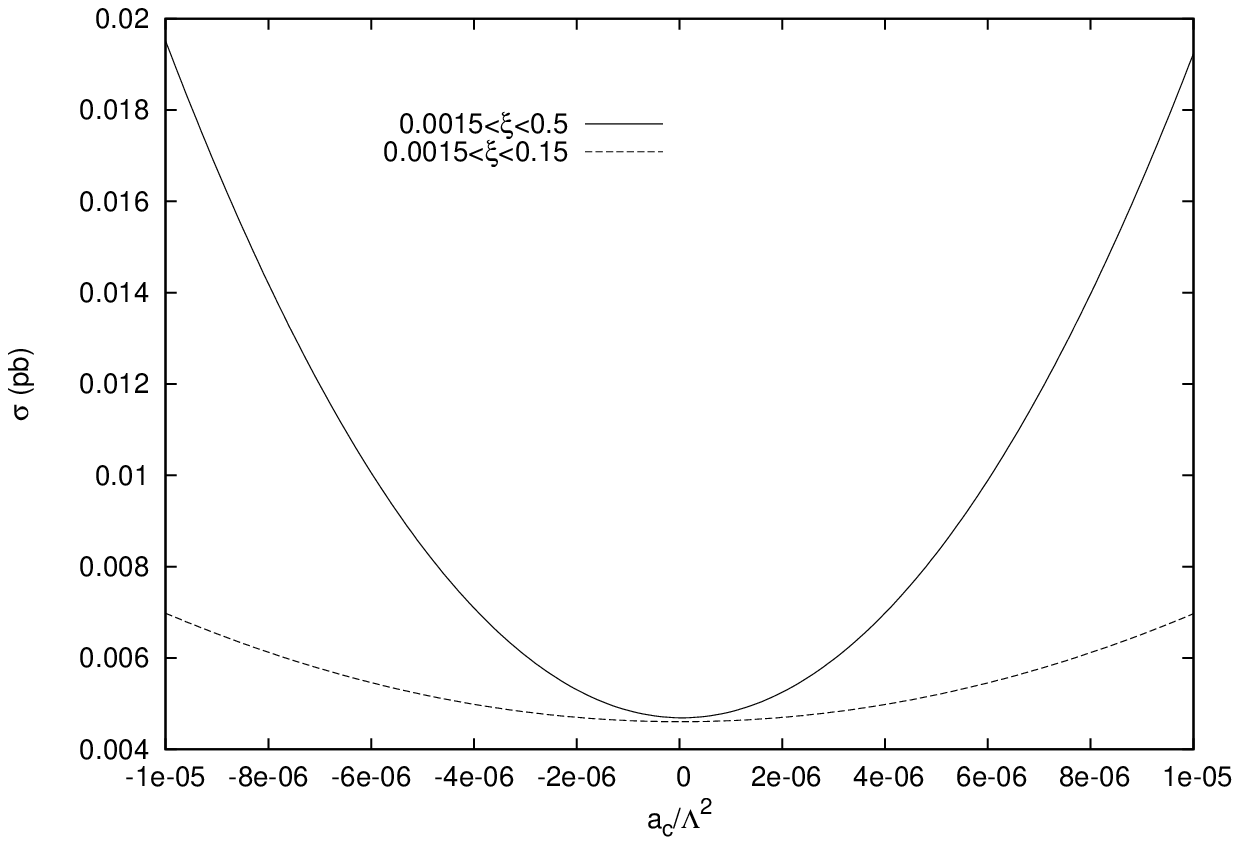}
\caption{Total cross section of $p p\to p\gamma p\to p \gamma q Z X$
as a function of anomalous coupling $\frac {a_{c}} {\Lambda^2}$ for
two different forward detector acceptances stated in the figure. The
center-of-mass energy of the proton-proton system is taken to be
$\sqrt s=14$ TeV. \label{fig4}}
\end{figure}

In Figs.\ref{fig3} and \ref{fig4} we plot the total cross section of
the process $p p\to p\gamma p\to p \gamma q Z X$ as a function of
anomalous couplings $\frac {a_{0}}{\Lambda^2}$ and $\frac
{a_{c}}{\Lambda^2}$ for the acceptances of $0.0015 <\xi < 0.5$ and
$0.0015 <\xi < 0.15$. We observe from these figures that cross
sections are large for $0.0015 <\xi < 0.5$ compared with $0.0015
<\xi < 0.15$ as expected. We also see from these figures that,
deviation of the anomalous cross section from its SM value is larger
for the coupling $\frac {a_{0}}{\Lambda^2}$ than $\frac
{a_{c}}{\Lambda^2}$. Therefore sensitivity limits on the coupling
$\frac {a_{0}}{\Lambda^2}$ are expected to be more restricted than
the limits on $\frac {a_{c}}{\Lambda^2}$.

In all results presented in this paper we assume that center-of-mass
energy of the proton-proton system is $\sqrt s=14$ TeV and cross
sections have evaluated numerically by a computer code GRACE
\cite{grace}.

\section{Limits on the anomalous couplings}

During statistical analysis we use two different method. We employ a
simple one-parameter $\chi^2$ test when the number of SM events is
greater than 10. On the other hand, we employ a Poisson distribution
when the number of SM events is less than or equal to 10. For AFP
and CMS-TOTEM scenarios SM cross sections for the main reaction $p
p\to p\gamma p\to p \gamma q Z X$ are 0.0046 pb and 0.0047 pb
respectively. Hence, the number of SM events exceeds 10 for
integrated luminosities which are equal or greater than
$100fb^{-1}$. Therefore for AFP and CMS-TOTEM scenarios we employ
both type of the statistical analysis depending on the luminosity.
Forward detectors have a capability to detect protons in a
continuous range of $\xi$. Therefore one can impose some cuts and
choose to work in a subinterval of the whole acceptance region.
Imposing such cuts on forward detector acceptance is useful in
suppressing the SM contribution. In addition to AFP and CMS-TOTEM
scenarios we will consider $0.1 <\xi < 0.15$ and $0.1 <\xi < 0.5$
subintervals of the whole AFP and CMS-TOTEM acceptance regions. For
these acceptances the number of SM events is less than 10. Therefore
it is very appropriate to set bounds on the couplings using a
Poisson distribution.

For the acceptances of $0.0015<\xi<0.5$ and $0.0015<\xi<0.15$ with a
high luminosity, $\chi^2$ analysis is performed. The $\chi^2$
function is defined by
\begin{eqnarray}
\chi^{2}=\left(\frac{\sigma_{SM}-\sigma_{AN}}{\sigma_{SM} \,\,
\delta}\right)^{2}
\end{eqnarray}
where $\sigma_{AN}$ is the cross section containing new physics
effects and $\delta=\frac{1}{\sqrt{N}}$ is the statistical error.
The expected number of events has been calculated considering the
leptonic decay channel of the Z boson as the signal $N=S\times
E\times \sigma_{SM} \times L_{int}\times BR(Z\to \ell \bar \ell)$,
where $\ell=e^-$ or $\mu^-$, $L_{int}$ is the integrated luminosity,
$E$ is the jet reconstruction efficiency and $S$ is the survival
probability factor. We have taken into account a jet reconstruction
efficiency of $E=0.6$ and survival probability factor of $S=0.7$.
This survival probability factor was proposed for the single W boson
photoproduction \cite{deFavereaudeJeneret:2009db,survivalfactor}.
 We assume that same survival factor is valid for our process.
ATLAS and CMS have central detectors with a pseudorapidity coverage
$|\eta|<2.5$. Therefore we place a cut of $|\eta|<2.5$ for final
state particles. Moreover, we also demand that the transverse
momenta of the final state photon and quark are greater than 15 GeV.
For the acceptances of $0.1 <\xi < 0.15$ and $0.1 <\xi < 0.5$, we
employ a Poisson distribution. Sensitivity limits are obtained
assuming the number of observed events equal to the SM prediction,
i.e., $N_{obs}=S\times E\times \sigma_{SM} \times L_{int}\times
BR(Z\to \ell \bar \ell)$. Upper limits of number of events $N_{up}$
at the 95\% C.L. can be calculated from the formula
\cite{deFavereaudeJeneret:2009db,Pierzchala}
\begin{eqnarray}
\sum_{k=0}^{N_{obs}}P_{Poisson}(N_{up};k)=0.05
\end{eqnarray}
Depending on the number of observed events, values for upper limits
$N_{up}$ can be found in Table 33.3 in Ref.\cite{Nakamura}. In Table
\ref{tab1} we present number of observed events and upper limits of
number of events for the cases in which the Poisson distribution has
been used. In Table \ref{tab1} the calculated $N_{obs}$ values are
rounded to the nearest integer. For instance, for forward detector
acceptance of $0.1<\xi<0.5$, $N_{obs}=0.69$ and 1.38 for
$L_{int}=100 fb^{-1}$ and $200 fb^{-1}$ respectively. Both of the
$N_{obs}$ values have been rounded to 1. The upper limits of number
of events $N_{up}$ can be directly converted to the limits of
anomalous couplings  $\frac {a_{0}}{\Lambda^2}$ and $\frac {a_{c}}
{\Lambda^2}$. In Tables \ref{tab2} and \ref{tab3}, we show 95\% C.L.
sensitivity limits on the anomalous couplings  $\frac
{a_{0}}{\Lambda^2}$ and $\frac {a_{c}} {\Lambda^2}$  for various
integrated luminosities and forward detector acceptances of
$0.0015<\xi<0.5$, $0.1<\xi<0.5$, $0.0015<\xi<0.15$ and
$0.1<\xi<0.15$. We see from Tables \ref{tab2} and \ref{tab3} that
our limits are at the order of $10^{-6}\;\textmd{GeV}^{-2}$ and
limits on the coupling $\frac {a_{0}}{\Lambda^2}$ are more
restricted than the limits on $\frac {a_{c}}{\Lambda^2}$.

In this paper we considered all tree-level SM contributions for the
subprocess $\gamma q\to \gamma q Z$ (Fig.\ref{fig2}). These
constitute major SM contributions. Any other SM contribution coming
to this subprocess is at the loop level and can be neglected
compared to tree-level SM contributions. The leading order
background process might be the pomeron exchange.  A pomeron emitted
from one proton beam can interact with the quarks of the other
proton and same final state can occur. But when we analyze in detail
we see that this background process is expected to has a minor
influence on sensitivity bounds. In the deep inelastic scattering
the virtuality of the struck quark is very high. During calculations
in this paper, the virtuality of the struck quark is taken to be
${Q}^2={m_Z}^2$ where $m_Z$ is the mass of the Z boson. Therefore,
when a pomeron strikes a quark it probably dissociates into partons.
These pomeron remnants can be detected by the calorimeters and
background from pomeron exchange can be eliminated. Furthermore,
survival probability for a pomeron exchange is considerably smaller
than that for a photon exchange. Hence, even if the background from
pomeron exchange can not be eliminated, it can not be much bigger
(probably smaller) than the tree-level SM contributions for the
photon exchange. Finally we would like to stress that our bounds are
not very sensitive to backgrounds. For instance, if we assume that
background cross section is 2 times bigger than the tree-level SM
contributions, our limits with a 200$fb^{-1}$ luminosity are spoiled
approximately a factor of 1.5 for $0.1<\xi<0.5$ and a factor of 1.3
for $0.0015<\xi<0.5$.

\section{Conclusions}
The process $p p\to p\gamma p\to p \gamma q Z X$ at the LHC with a
center-of-mass energy of 14 TeV probes anomalous quartic
$ZZ\gamma\gamma$ couplings with a far better sensitivity than the
current experimental bounds. It allows to improve the current
sensitivity by almost four orders of magnitude. The potential of LHC
to probe anomalous quartic $ZZ\gamma\gamma$ couplings was examined
via weak boson fusion processes $qq \to qq\gamma\gamma$ and $qq \to
qq\gamma Z(\to l^{+}l^{-})$ \cite{lietti} and photon-photon fusion
process $pp\to p \gamma \gamma p\to p Z Z p$
\cite{lhc2,lhc11,deFavereaudeJeneret:2009db,Pierzchala}. In papers
\cite{deFavereaudeJeneret:2009db,Pierzchala} authors considered
semi-leptonic decay channel of the final Z bosons ,i.e., $ZZ\to
\ell^+ \ell^- jj$ where $j$ refers to jets. On the other hand in
papers \cite{lhc2,lhc11} authors considered fully-leptonic decay
channel of the final Z bosons.  The bounds obtained in
\cite{lhc2,lhc11} are considerably weaker than the bounds obtained
in papers \cite{deFavereaudeJeneret:2009db,Pierzchala}. This
probably originates from the fact that semi-leptonic decay channel
of the final Z bosons has a large branching ratio compared to
fully-leptonic decay channel. In our paper we have considered
leptonic decay channel of final Z. Therefore it is more appropriate
to compare our bounds with the bounds obtained in \cite{lhc2,lhc11}.
The bounds obtained in \cite{lietti,lhc2,lhc11} are of the same
order as our bounds. Anomalous quartic $ZZ\gamma\gamma$ couplings
were also studied for future International Linear Collider (ILC) and
its operating modes of $e\gamma$ and $\gamma \gamma$. The limits
expected to be obtained for such a machine are comparable with the
LHC bounds \cite{Atag:2007ct}.

\appendix*
\section{Equivalent Photon Spectrum}
Taking into consideration the electromagnetic form factors of the
proton, equivalent photon spectrum of virtuality $Q^2$ and energy
$E_\gamma$ is given by the following formula \cite{budnev,Baur,fd18}
\begin{eqnarray}
\label{spectrum1}
\frac{dN_\gamma}{dE_{\gamma}dQ^{2}}=\frac{\alpha}{\pi}\frac{1}{E_{\gamma}Q^{2}}
[(1-\frac{E_{\gamma}}{E})
(1-\frac{Q^{2}_{min}}{Q^{2}})F_{E}+\frac{E^{2}_{\gamma}}{2E^{2}}F_{M}]
\end{eqnarray}
where
\begin{eqnarray}
&&Q^{2}_{min}=\frac{m^{2}_{p}E^{2}_{\gamma}}{E(E-E_{\gamma})},
\;\;\;\; F_{E}=\frac{4m^{2}_{p}G^{2}_{E}+Q^{2}G^{2}_{M}}
{4m^{2}_{p}+Q^{2}} \\
G^{2}_{E}=&&\frac{G^{2}_{M}}{\mu^{2}_{p}}=(1+\frac{Q^{2}}{Q^{2}_{0}})^{-4},
\;\;\; F_{M}=G^{2}_{M}, \;\;\; Q^{2}_{0}=0.71 \mbox{GeV}^{2}
\end{eqnarray}
In the above formula, E is the energy of the incoming proton beam
and $m_{p}$ is the mass of the proton. $F_{E}$ and $F_{M}$ are
functions of the electric and magnetic form factors. $\mu^{2}_{p}$
is the magnetic moment of the proton. It is taken to be
$\mu^{2}_{p}=7.78$. $dQ^2$ integration in (\ref{spectrum1}) can be
easily performed analytically. After integration over $Q^2$,
(\ref{spectrum1}) takes the form of \cite{fd11}
\begin{eqnarray}
\label{spectrum2} \frac{dN_\gamma}{dE_{\gamma}}=\frac{\alpha}{\pi
E_{\gamma}}
\left(1-\frac{E_{\gamma}}{E}\right)\left[\varphi\left(\frac{Q^{2}_{max}}{Q^{2}_0}\right)
-\varphi\left(\frac{Q^{2}_{min}}{Q^{2}_0}\right)\right]
\end{eqnarray}
where the function $\varphi$ is defined by
\begin{eqnarray}
\varphi(x)=(1+ay)\left[-ln(1+\frac{1}{x})+\sum_{k=1}^3\frac{1}{k(1+x)^k}\right]+\frac{y(1-b)}{4x(1+x)^3}\nonumber\\
+c\left(1+\frac{y}{4}\right)\left[ln\left(\frac{1-b+x}{1+x}\right)+\sum_{k=1}^3\frac{b^k}{k(1+x)^k}\right]
\end{eqnarray}
where
\begin{eqnarray}
y=\frac{E_\gamma^2}{E(E-E_\gamma)},\;\;\;\;\;\;a=\frac{1+\mu^{2}_{p}}{4}+\frac{4m^2_{p}}{Q^{2}_0}\approx7.16 \nonumber\\
b=1-\frac{4m^2_{p}}{Q^{2}_0}\approx-3.96,\;\;\;\;\;\;c=\frac{\mu^{2}_{p}-1}{b^4}\approx0.028
\end{eqnarray}
Here $Q^{2}_{max}$ and $Q^{2}_{min}$ are the upper and lower bounds
of the integration. The contribution to the integral above
$Q^{2}_{max}\approx2\;GeV^2$ is negligible. Therefore during
calculations we set $Q^{2}_{max}=2\;GeV^2$.


\newpage

\begin{table}
\caption{The number of observed events $N_{obs}$ and corresponding
values for upper limits $N_{up}$ at 95\% C.L. The calculated
$N_{obs}$ values are rounded to the nearest integer. \label{tab1}}
\begin{tabular}{ccccc}
\hline \hline
& $0.0015<\xi<0.5$ & &  $0.0015<\xi<0.15$\\

 $L(fb^{-1})$ \ \ \ & $N_{obs}$ \ \ \ \ \ \ \ \ \ $N_{up}$ & \ \ \ \ \ \ \ \ \ \ & $N_{obs}$ \ \ \ \ \ \ \ \ \ $N_{up}$ \\
 \hline
  30  &  4 \ \ \ \ \ \ \ \ \ \ \ \ \ 9.15  &  & 4 \ \ \ \ \ \ \ \ \ \ \ 9.15     \\
  50  &  7 \ \ \ \ \ \ \ \ \ \ \ \  13.15   &  & 7  \ \ \ \ \ \ \ \ \ \ 13.15  \\
\hline
& $0.1<\xi<0.5$ & &  $0.1<\xi<0.15$\\

 $L(fb^{-1})$ \ \ \ & $N_{obs}$ \ \ \ \ \ \ \ \ \ $N_{up}$ & \ \ \ \ \ \ \ \ \ \ & $N_{obs}$ \ \ \ \ \ \ \ \ \ $N_{up}$ \\
 \hline
  30  &  0 \ \ \ \ \ \ \ \ \ \ \ \ \ 3.00  &  & 0 \ \ \ \ \ \ \ \ \ \ \  3.00 \\
  50  &  0 \ \ \ \ \ \ \ \ \ \ \ \ \ 3.00 &  & 0  \ \ \ \ \ \ \ \ \ \ \ 3.00 \\
 100  &  1 \ \ \ \ \ \ \ \ \ \ \ \ \ 4.74  &  & 1  \ \ \ \ \ \ \ \ \ \ \ 4.74     \\
 200  &  1 \ \ \ \ \ \ \ \ \ \ \ \ \ 4.74   &  & 1  \ \ \ \ \ \ \ \ \ \ \ 4.74  \\
  \hline \hline
\end{tabular}
\end{table}

\begin{table}
\caption{95\% C.L. sensitivity bounds of the couplings $\frac
{a_{0}}{\Lambda^2}$ and $\frac {a_{c}} {\Lambda^2}$ for various LHC
luminosities and forward detector acceptances of $0.0015<\xi<0.5$
and $0.1<\xi<0.5$. The center of mass energy of the proton-proton
system is taken to be $ \sqrt{s}=14 $ TeV.\label{tab2}}
\begin{ruledtabular}
\begin{tabular}{ccccccc}
& $0.0015<\xi<0.5$ & $0.1<\xi<0.5$\\

 $L(fb^{-1})$ & $\frac{a_{0}}{\Lambda^2}(\times 10^{-6}\;\textmd{GeV}^{-2})$\;\;  $\frac{a_{c}}{\Lambda^2}(\times 10^{-6}\;\textmd{GeV}^{-2})$ &
 $\frac{a_{0}}{\Lambda^2}(\times 10^{-6}\;\textmd{GeV}^{-2})$\ \;\;  $\frac{a_{c}}{\Lambda^2}(\times 10^{-6}\;\textmd{GeV}^{-2})$\\
 \hline
  30  & -4.0;4.0 \ \ \ \ \ \ \ \ \ \ \ -6.4;6.5   & -2.8;2.8  \ \ \ \ \ \ \ \ \ \ \ -4.1;4.1   \\
  50  & -3.4;3.4 \ \ \ \ \ \ \ \ \ \ \ -5.6;5.7   & -2.2;2.2  \ \ \ \ \ \ \ \ \ \ \ -3.1;3.1 \\
  100 & -2.5;2.5 \ \ \ \ \ \ \ \ \ \ \ -4.1;4.2   & -1.9;1.9  \ \ \ \ \ \ \ \ \ \ \ -2.7;2.7 \\
  200 & -2.1;2.1 \ \ \ \ \ \ \ \ \ \ \ -3.4;3.5   & -1.2;1.2  \ \ \ \ \ \ \ \ \ \ \ -1.7;1.7  \\
\end{tabular}
\end{ruledtabular}
\end{table}

\begin{table}
\caption{95\% C.L. sensitivity bounds of the couplings $\frac
{a_{0}}{\Lambda^2}$ and $\frac {a_{c}}{\Lambda^2}$ for various LHC
luminosities and forward detector acceptances of $0.0015<\xi<0.15$
and $0.1<\xi<0.15$. The center of mass energy of the proton-proton
system is taken to be $\sqrt{s}=14 $ TeV.\label{tab3}}
\begin{ruledtabular}
\begin{tabular}{ccccccc}
& $0.0015<\xi<0.15$ & $0.1<\xi<0.15$\\

 $L(fb^{-1})$ & $\frac{a_{0}}{\Lambda^2}(\times 10^{-6}\;\textmd{GeV}^{-2})$\;\;  $\frac{a_{c}}{\Lambda^2}(\times 10^{-6}\;\textmd{GeV}^{-2})$ &
 $\frac{a_{0}}{\Lambda^2}(\times 10^{-6}\;\textmd{GeV}^{-2})$\ \;\;  $\frac{a_{c}}{\Lambda^2}(\times 10^{-6}\;\textmd{GeV}^{-2})$\\
 \hline
  30  & -10.9;11.0 \ \ \ \ \ \ \ \ \ \ \ -16.2;16.2   &  -9.5;9.5 \ \ \ \ \ \ \ \ \ \ \ -13.7;13.7   \\
  50  & -9.5;9.6 \ \ \ \ \ \ \ \ \ \ \ -14.2;14.2   &  -7.3;7.3 \ \ \ \ \ \ \ \ \ \ \ -10.5;10.5   \\
  100 & -6.9;7.0 \ \ \ \ \ \ \ \ \ \ \ -10.3;10.3   &  -6.4;6.4 \ \ \ \ \ \ \ \ \ \ \ -9.2;9.2   \\
  200 & -5.8;5.9 \ \ \ \ \ \ \ \ \ \ \ -8.6;8.7     &  -4.3;4.3 \ \ \ \ \ \ \ \ \ \ \ -6.1;6.1   \\
\end{tabular}
\end{ruledtabular}
\end{table}


\begin{thebibliography}{99}

\bibitem{boudjema1} G. Belanger and F. Boudjema Phys. Lett. B 288, 201 (1992).

\bibitem{boudjema2} G. Belanger and F. Boudjema Phys. Lett. B 288, 210 (1992).

\bibitem{eboli1} O. J. P. Eboli, M. C. Gonzalez-Garcia and S. F. Novaes,
Nucl. Phys. {\bf B411}, 381 (1994).

\bibitem{Abbiendi:2004bf}
  G.~Abbiendi {\it et al.}  [OPAL Collaboration],
  Phys.\ Rev.\ D {\bf 70}, 032005 (2004)  [hep-ex/0402021].  

\bibitem{cdf1} A. Abulencia {\it et al.} (CDF Collaboration), Phys. Rev. Lett. {\bf 98}, 112001
(2007); arXiv:hep-ex/0611040.

\bibitem{cdf4} T. Aaltonen {\it et al.} (CDF Collaboration), Phys. Rev. Lett. {\bf 102}, 222002
(2009); arXiv:0902.2816 [hep-ex].

\bibitem{cdf3} T. Aaltonen {\it et al.} (CDF Collaboration), Phys. Rev. Lett. {\bf 102}, 242001
(2009); arXiv:0902.1271 [hep-ex].


\bibitem{fd11}O. Kepka and C. Royon, Phys. Rev. D {\bf 78}, 073005 (2008);
arXiv:0808.0322 [hep-ph].

\bibitem{fd12} V.A. Khoze, A.D. Martin and M.G. Ryskin,
Eur. Phys. J. C {\bf 23}, 311 (2002); arXiv:hep-ph/0111078.

\bibitem{fd13}N. Schul and K. Piotrzkowski, Nucl. Phys. B, Proc.
Suppl., {\bf 179}, 289 (2008); arXiv:0806.1097 [hep-ph].


\bibitem{lhc1}S. M. Lietti, A. A. Natale, C. G. Roldao and R.
Rosenfeld, Phys. Lett. B {\bf 497}, 243 (2001);
arXiv:hep-ph/0009289.

\bibitem{lhc2}E. Chapon, C. Royon and O. Kepka , Phys. Rev. D {\bf 81}, 074003
(2010); arXiv:0912.5161 [hep-ph].

\bibitem{lhc3} S. Ata\u{g}, S. C. \.{I}nan and \.{I}. \c{S}ahin, Phys. Rev. D {\bf 80}, 075009 (2009);
 arXiv:0904.2687 [hep-ph].

\bibitem{lhc4} \.{I}. \c{S}ahin and S. C. \.{I}nan, JHEP {\bf 09}, 069
(2009); arXiv:0907.3290 [hep-ph].

\bibitem{lhc5} S. Ata\u{g}, S. C. \.{I}nan and \.{I}. \c{S}ahin, JHEP {\bf
09}, 042 (2010);  arXiv:1005.4792 [hep-ph].

\bibitem{lhc6} S. C. \.{I}nan, Phys. Rev. D {\bf 81}, 115002
(2010); arXiv:1005.3432 [hep-ph].

\bibitem{lhc7} S. Ata\u{g} and A. A. Billur, JHEP {\bf 11}, 060 (2010); arXiv:1005.2841
[hep-ph].

\bibitem{lhc8} M.G. Albrow, T.D. Coughlin and J.R. Forshaw, Prog. Part. Nucl. Phys. {\bf
65},149-184 (2010); arXiv:1006.1289 [hep-ph].

\bibitem{lhc9}  \.{I}. \c{S}ahin, and A. A. Billur, Phys. Rev. D {\bf
83}, 035011 (2011); arXiv:1101.4998 [hep-ph].

\bibitem{lhc10}  \.{I}. \c{S}ahin, and M. Koksal, JHEP {\bf 03}, 100 (2011); arXiv:1010.3434 [hep-ph].


\bibitem{lhc11} R.~S.~Gupta, Phys.\ Rev.\ D {\bf 85}, 014006 (2012)
[arXiv:1111.3354 [hep-ph]].

\bibitem{lhc12}  \.{I}. \c{S}ahin, Phys.\ Rev.\ D {\bf 85}, 033002 (2012)  [arXiv:1201.4364 [hep-ph]].


\bibitem{budnev} V. M. Budnev, I. F. Ginzburg, G. V. Meledin and V.
G. Serbo, Phys. Rep. {\bf 15}, 181 (1975).

\bibitem{Baur} G. Baur {\it et al.}, Phys. Rep. {\bf 364}, 359
(2002).

\bibitem{fd18}K. Piotrzkowski, Phys. Rev. D {\bf 63}, 071502 (2001) [hep-ex/0009065].

\bibitem{rouby} X. Rouby, Ph.D. thesis, Universite catholique de Louvain [UCL-Thesis
135-2008, CMS TS-2009/004], 2008.

\bibitem{deFavereaudeJeneret:2009db}
  J.~de Favereau de Jeneret, V.~Lemaitre, Y.~Liu, S.~Ovyn, T.~Pierzchala, K.~Piotrzkowski, X.~Rouby, N.~Schul and M. Vander Donckt,
  arXiv:0908.2020 [hep-ph].

\bibitem{royon}  C. Royon {\it et al.}  (RP220 Collaboration), arXiv:0706.1796 [physics.ins-det], {\it Proceedings for the DIS 2007 workshop, Munich,
2007}.

\bibitem{albrow} M.G. Albrow {\it et al.} (FP420 R and D
Collaboration), JINST {\bf 4}, T10001 (2009); arXiv:0806.0302
[hep-ex].

\bibitem{avati} V. Avati and K. Osterberg, Report No.
CERN-TOTEM-NOTE-2005-002, 2006.

\bibitem{pdf}  A. D. Martin, W. J. Stirling, R. S. Thorne and G.
Watt, Phys. Lett. B {\bf 652}, 292 (2007); arXiv:0706.0459 [hep-ph].

\bibitem{grace} T. Kaneko in {\it New Computing Techniques in Physics
Research}, edited by D. Perret-Gallix, W. Wojcik  (CNRS, Paris,
1990); MINAMI-TATEYA Group, KEK Report No. 92-19, 1993; F. Yuasa
{\it et al.}, Prog. Theor. Phys. Suppl. {\bf 138}, 18 (2000).


\bibitem{survivalfactor} V.A. Khoze, A.D. Martin and M.G. Ryskin,  Eur. Phys. J. C  {\bf24}, 459 (2002).

\bibitem{Pierzchala} T. Pierzchala and K. Piotrzkowski, Nucl.\ Phys.\ Proc.\ Suppl.\
{\bf 179-180}, 257 (2008) arXiv:0807.1121 [hep-ph].

\bibitem{Nakamura} K. Nakamura et al. (Particle Data Group), J. Phys. G 37, 075021
(2010).

\bibitem{lietti} O. J. P. Eboli, M. C. Gonzalez-Garcia and S. M.
Lietti, Phys. Rev. D {\bf 69}, 095005 (2004) [hep-ph/0310141].


\bibitem{Atag:2007ct}
  S.~Ata\u{g} and \.{I}. \c{S}ahin,
   Phys.\ Rev.\ D {\bf 75}, 073003 (2007)  [hep-ph/0703201 [HEP-PH]].



\end{thebibliography}
\end{document}